\documentclass[aps,prb,preprint,nofootinbib]{revtex4-1}
\usepackage{float}
\usepackage{amsfonts}
\usepackage{amsmath}
\usepackage{amssymb}
\usepackage{graphicx}%
\usepackage{graphics}
\usepackage{color}
\usepackage{notes2bib}
\usepackage{pstricks}
\begin{document}

\begin{center}
\noindent{\textbf{{\LARGE {{Exact solutions for a ferromagnet with Dzyaloshinskii-Moriya interaction} }}}}

\smallskip\smallskip

\smallskip\smallskip

\smallskip\smallskip

\smallskip\smallskip
 
\noindent{{Nicol\'as Grandi$^{a}$, Marcela Lagos$^b$, Julio Oliva$^{c}$, Aldo Vera$^{b}$}}

\smallskip\smallskip

\smallskip\smallskip

$^{a}$\textit{{Instituto de F\'{\i}sica de La Plata - CONICET \& Departamento de
F\'{\i}sica - UNLP,}\newline\textit{C.C. 67, 1900 La Plata, Argentina.}} \\
$^{b}$\textit{Instituto de Ciencias F\'isicas y Matem\'aticas, Universidad Austral de Chile,\\ Casilla 567, Valdivia, Chile.}\\
$^{c}$\textit{\ Departamento de F\'isica, Universidad de Concepci\'on,
Casilla 160-C, Concepci\'on, Chile.}
\\{\small grandi@fisica.unlp.edu.ar, marcelagosf@gmail.com, juoliva@udec.cl, aldo.vera@uach.cl}


\end{center}

\smallskip\smallskip

\smallskip\smallskip

\smallskip\smallskip

\smallskip

\begin{center}
\textbf{Abstract}
\end{center}

On the two-dimensional non-linear $\Sigma$-model describing a ferromagnet with Dzyaloshinskii-Moriya interaction, we build three families of exact static solutions depending on a single Cartesian variable. One of them describes a clockwise helix configuration, that characterizes the ground state of the system. A second one corresponds to a counterclockwise helix, representing an excited state. These two families of solutions are parameterized by a continuous parameter that depends on the magnetic field and the Dzyaloshinskii-Moriya coupling. Finally, the third family exists only for isolated values of the same parameter, corresponding to a discrete family of Viviani curves on the target sphere. The degeneracy of the resulting spectrum suggests that an approximate symmetry may emerge at specific values of the magnetic field, at which additional solutions could then exist. 
\newpage
\section{Introduction}
Spin textures have been a subject of intense scrutiny in the last years. This is partially due to the realization that localized spin structures with non-trivial topological properties can be globally stabilized by a Dzyaloshinskii-Moriya interaction\cite{Dzyaloshinskii1964}. In particular, configurations known as {\em magnetic Skyrmions} have been theoretically predicted\cite{Yablonskii1989, Bogdanov1994, Bogdanov1999} and experimentally found \cite{Muhlbauer2009, Ishikawa1984, Lebech1995, Shibata2013, Lebech1989, Uchida2008, Yu2011, Wilhelm2011, Beille1983, Grigoriev2007, Grigoriev2009, Onose2005, Kezsmarki2015, Seki2012, Adams2012} in materials such as 
MnSi, Mn$_{1-x}$Fe$_x$Ge, FeGe, Fe$_{1-x}$Co$_x$Si, GaV$_4$S$_8$ and Cu$_2$SeO$_3$.

When the Skyrmion density is high enough, Skyrmions get packed into a {\em Skyrmion lattice} structure. Such non-ionic lattice has non-standard phononic degrees of freedom, that could mediate High-$T_c$ superconductivity without an isotopic effect \cite{Baskaran2011a}. The Skyrmion lattice is stable at an intermediate range of magnetic fields at sufficiently low temperatures, degenerating into the ferromagnetic phase at high magnetic fields, and into a configuration depending on a single spatial direction known as {\em helix} at low magnetic fields \cite{Bogdanov1994,Yi2009}. 

The dynamics of a ferromagnetic spin system is modeled in the continuum limit with a non-linear $\Sigma$-model. The Dzyaloshinskii-Moriya interaction shows up as a parity breaking term coupling spatial and internal directions. The phases are generically studied in terms of approximate solutions by means of a variational approach, both for the Skyrmion lattice phase and the helix phase\cite{Han2010,Grandi2018a}. Even if the helix phase can be treated exactly \cite{Dzyaloshinskii1964,Nogueira2018}, no analytic solution is known at present for the Skyrmion lattice configuration. 

In the present note we explore exact solutions of the non-linear $\Sigma$-model with a Dzyal\-oshin\-skii-Moriya term, concentrating in solutions with non-trivial topology depending on a single spatial dimension. Along with the previously known helix solutions, we find an additional family that we call {\em Viviani} solutions. They exist at isolated values of the magnetic field, and result into a very interesting structure of the energy spectrum. 

The paper is organized as follows: in Section \ref{sec:model} we write the Hamiltonian of the model to be minimized by the classical solutions and define a useful parameterization of the dynamical variables. In Section \ref{sec:helix} we review the known exact helix solutions of the equations of motion. In Section \ref{sec:viviani} we present the new solutions and study their properties. Finally, in Section \ref{sec:discussion} we analyze the resulting energy spectrum and discuss its implications. 

\newpage

\section{The non-linear $\Sigma$-model and the field equations}
\label{sec:model}
The Hamiltonian for a ferromagnetic system with Dzyaloshinskii-Moriya interaction in the continuum limit reads
\begin{equation}
H=\frac{1}{2}\int d^{2}\underline{x}\left(  J\ \nabla_{\underline{i}}n_{a}\nabla_{\underline{i}}n_{a}%
+2D\epsilon^{a\underline{i}b}n_{a}\nabla_{\underline{i}}n_{b}-2B^{a}n_{a}-\lambda\left(n_an_a-1\right)\right)  \ ,
\end{equation}
where $J$ is the exchange contribution, $D$ is the Dzyaloshinskii-Moriya coupling and $B^a$ is an external magnetic field
that we specialize to the $\underline{z}$-direction $B^a=\delta^{\underline{z}a} B$. Here $\lambda$ is a Lagrange multiplier ensuring that the field $n_a$ lives in a target unitary sphere.

For non-vanishing $B$, we rewrite the above Hamiltonian in a simpler form using the re-scalings $x_{\underline{i}}=2D x_{{i}}/B$, obtaining
\begin{equation}
H=\frac{J}{2}\int d^{{2}}x
\left(  \nabla_{{i}}n_{a}\nabla_{{i}}n_{a}
+p\left(\epsilon^{a{i}b}n_{a}\nabla_{{i}}n_{b}-2\,\delta^{za}n_{a}\right)
-\lambda\left(n_an_a-1\right)
\right)  \ ,
\end{equation}
where $p=4D^2/{BJ}$ is the only parameter characterizing the model, and the Lagrange multiplier $\lambda$ has been re-scaled accordingly. The resulting field equations read
\begin{eqnarray}
&&\nabla^2 n_a-p\left( \epsilon^{a\underline{i}b}\nabla_{{i}}n_b
-\delta^{za}\right)+\lambda n_a=0\,,
\nonumber\\
&&n_a n_a -1=0\,.
\end{eqnarray}
We will solve the above set of equations in a square box in the $x^{{i}}=(x,y)$ plane with sides $(L_x, L_y)$, imposing periodic boundary conditions.

We are interested in configurations depending on a single Cartesian variable that we choose as $x$. In other words, we particularize the equations of motion to the Ansatz $n_a=n_a(x)$, obtaining
\begin{eqnarray}
&&n''_a+p\left( \epsilon^{xab}n'_b
+  \delta^{za}\right)+\lambda n_a=0\,,
\nonumber\\
&&n_a n_a -1=0\,,
\end{eqnarray}
where a prime $(')$ denotes derivative with respect to $x$. These equations can be written explicitly in components, as
\begin{eqnarray}
&&n''_x+\lambda n_x=0\,,
\nonumber\\
&&n''_y+p n'_z
+\lambda n_y=0\,,
\nonumber\\
&&n''_z-p n'_y+ p+\lambda n_z=0\,,
\nonumber\\
&&n_x n_x+n_y n_y+n_z n_z -1=0\,.
\end{eqnarray}
The above expressions suggest a particularly useful parameterization of the fields, defined as $n_x=X$ and ${\mathcal{Z}}=n_y+in_z$. In terms of it, the equations read
\begin{eqnarray}
&&X''+\lambda X=0\,,
\nonumber\\
&&\mathcal{Z}''-i p\mathcal{Z}' +
\lambda \mathcal{Z} + i p=0\,,
\nonumber\\
&&X^2+ \mathcal{Z}\mathcal{Z}^*-1=0\,.
\end{eqnarray}
In what follows, we will search for solutions $X(x),\mathcal{Z}(x)$ of these equations, with the conditions
\begin{eqnarray}
X(L_x)&=&X(0)\,,
\\
\mathcal{Z}(L_x)&=&\mathcal{Z}(0)\,,
\end{eqnarray}
and where $\lambda$ and $X$ are reals. 

\vspace{1cm}
In order to compare the energies of the different solutions, we need an expression for the on-shell form of the Hamiltonian.  
The Hamiltonian evaluated on configurations depending on a single Cartesian variable reads
\begin{equation}
H =\frac{JL_y}{2}\int_0^{L_x} dx
\left(  n'_{a}n'_{a}
+p\left(-\epsilon^{xab}n_{a}n'_{b}-2\,\delta^{za}n_{a}\right)
-\lambda\left(n_an_a-1\right)
\right)  \ .
\label{eq:reduced}
\end{equation}
Integrating by parts in the first term and using the equations of motion, we get to
\begin{equation}
H^{\sf on-shell}=\frac{JL_y}{2}
\int_0^{L_x} dx
\left(
\lambda
-p\,\delta^{za}n_{a}
\right)  \ ,
\end{equation}
or, in terms of our parameterization 
\begin{equation}
H^{\sf on-shell}=\frac{JL_y}{2}
\int_0^{L_x} dx
\left(
\lambda
-p \,{\rm Im}\left[\mathcal{Z}\right]
\right)  \ .
\label{eq:honshell}
\end{equation}

There is an interesting analogue of the reduced Hamiltonian
in terms of a charged particle under an electric and a magnetic field, constrained to move on a $S^2$. Indeed, such dynamics is described by the following action
\begin{equation}
 S=\frac{1}{2}\int dt\biggl((\dot{\vec{r}})^2
 +(\vec{r}\times\dot{\vec r})\cdot{\vec{B}}+2\vec{E}\cdot\vec{r}-{\lambda}(\vec{r}\cdot\vec{r}-1)\biggl)\ ,
\end{equation}
that can be mapped into \eqref{eq:reduced} making the identifications
\begin{equation}
\vec r(t)=\vec n(x)\,,\qquad
 \vec{E}=-p \check k\ ,\qquad \vec{B}=-p \check{i} \ . 
\end{equation}

\subsection{Helix solutions}
\label{sec:helix}
We first look for solutions located at a meridian of the internal sphere perpendicular to the $n_x$ direction. In other words, we use the Ansatz
\begin{equation}
X=0\,.
\end{equation}
As it develops in the $x$ direction, the extreme of the vector $n^a$ traces a spring form that we call the {\em helix}. Plugging the Ansatz into the equations of motion, they become
\begin{eqnarray}
&&\mathcal{Z}''-i p\mathcal{Z}' +
\lambda \mathcal{Z} + i p=0\,,
\nonumber\\
&&\mathcal{Z}\mathcal{Z}^*-1=0\,.
\end{eqnarray}
The second equation implies that $\mathcal{Z}$ is a phase, while the first can be multiplied by $\mathcal{Z}^*$ and then easily solved for $\lambda$. We get
\begin{eqnarray}
&&\lambda=-(\mathcal{Z}''-i p\mathcal{Z}' +i p)\mathcal{Z}^*\,,
\nonumber\\
&&\mathcal{Z}=e^{i \varphi(x)}\,.
\end{eqnarray}
This is a valid solution of the equations for any function $\varphi(x)$ but, in our case, we need also to ensure that 
the Lagrange multiplier is real, which imposes a condition on $\varphi(x)$ via
\begin{equation}
{\rm Im}\lambda= \varphi''+p\cos\varphi=0\,.
\end{equation}
This is the pendulum equation, that is solved in terms of elliptic functions
\begin{equation}
\varphi= \frac{\pi}{2}\pm 2\,{\rm Am}\!\left[\sqrt{\frac{p}{p_0}}(x-x_0), -p_0\right]\,,
\end{equation}
where ${\rm Am}[\cdot,\cdot]$ represents the Jacobi Amplitude, and $p_0$ and $x_0$ are integration constants. The $\pm$ sign determines the rotation of the vector $n^a$ around the $x$ axis as we move along it, the $-$ sign corresponding to a clockwise helix. 

The resulting function $\mathcal{Z}$ is periodic, with period $2 K(-p_0)\sqrt{p_0/p}$, where $K(\cdot)$ is the Complete Elliptic Integral of the First Kind. This implies that the integration constant $p_0$ must satisfy
\begin{equation}
p_0 K^2\!\left(-p_0\right)
=
\frac{p\, L_x^2}{4 m^2}\,,
\label{eq:K}
\end{equation}
where $m\in\mathbb{Z}$ is an integer number representing the winding of the resulting helix around the $x$-axis. Notice that this makes $p_0$ a function of $m$ and $p$.

\begin{figure}[H]
~~~~
\includegraphics[height=8cm]{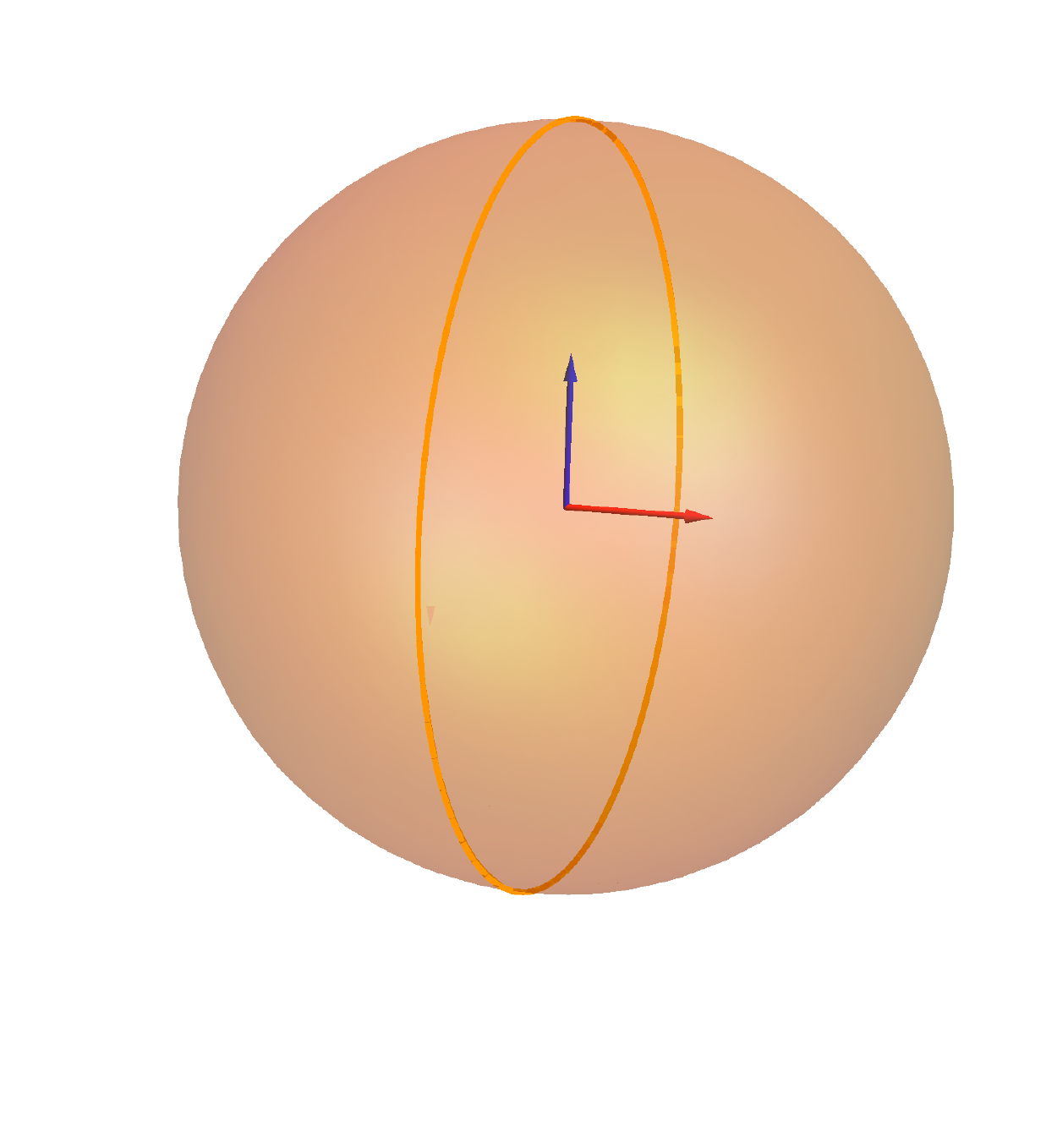}
~~~~~~~~
\includegraphics[height=8cm]{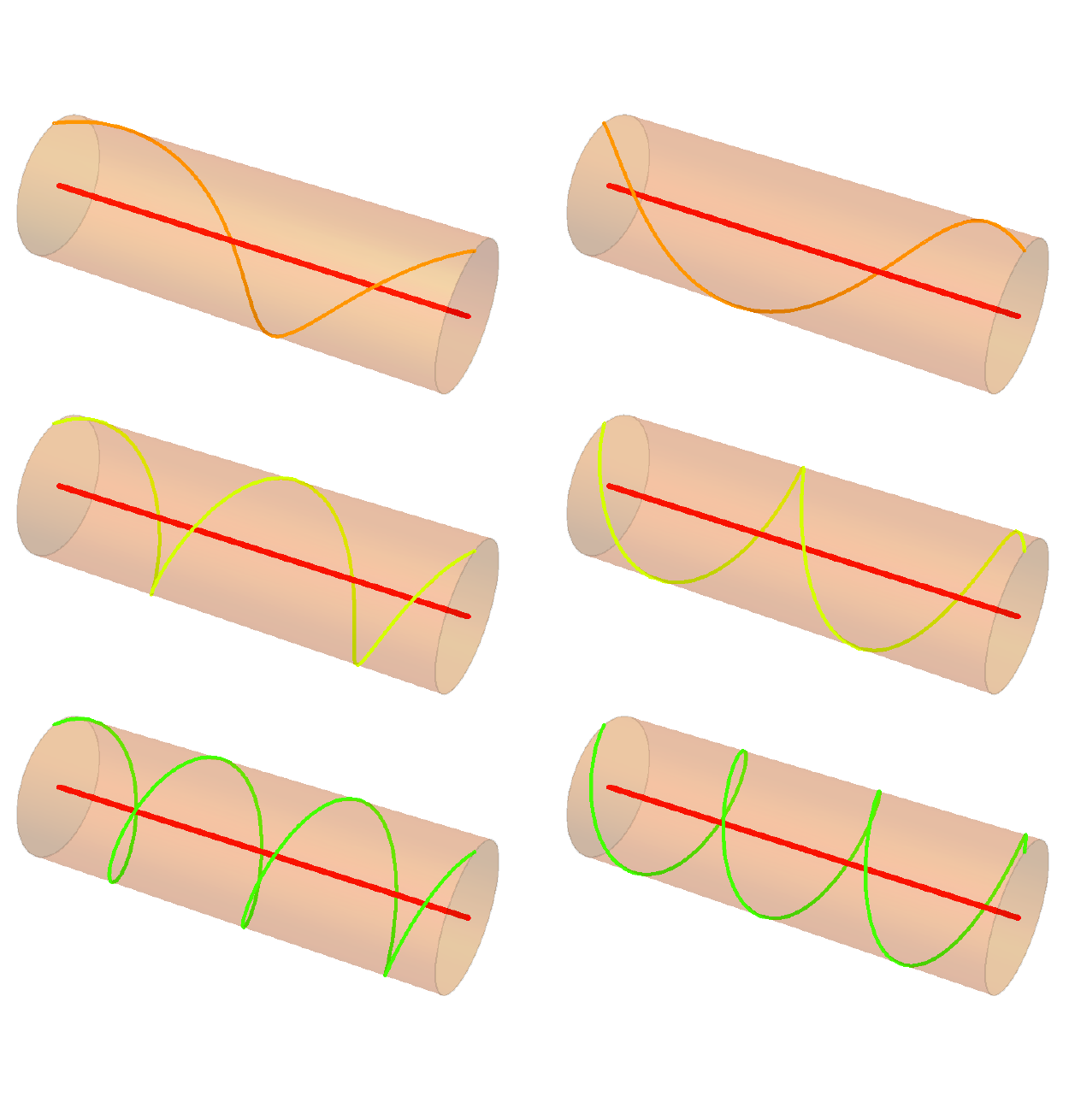}
\vspace{-1cm}
\caption{The helix solutions. It plots a circle at a meridian in the target sphere (left), which is on the plane of the magnetic field (blue arrow) and perpendicular to the helix axis (red arrow). As it develops in real space as a function of $x$ (right), it turns around the $x$-axis (red line) either clockwise when the solution with a $-$ sign is chosen (left column) or counterclockwise when the solution with a $+$ sign is chosen (right column). The number $m$ determines the winding, being $1,2,3$ from top to bottom.
\label{helix}}
\end{figure}

Plots of the resulting helix solutions  with both chiralities and for different windings 
are shown in Fig.\ref{helix}. They can be understood in terms of the electromagnetic analog proposed in section \ref{sec:model}: the combined analog electric and magnetic forces allow for a stable motion along the $X=0$ meridian. 

\begin{figure}[H]
\includegraphics[height=8cm]{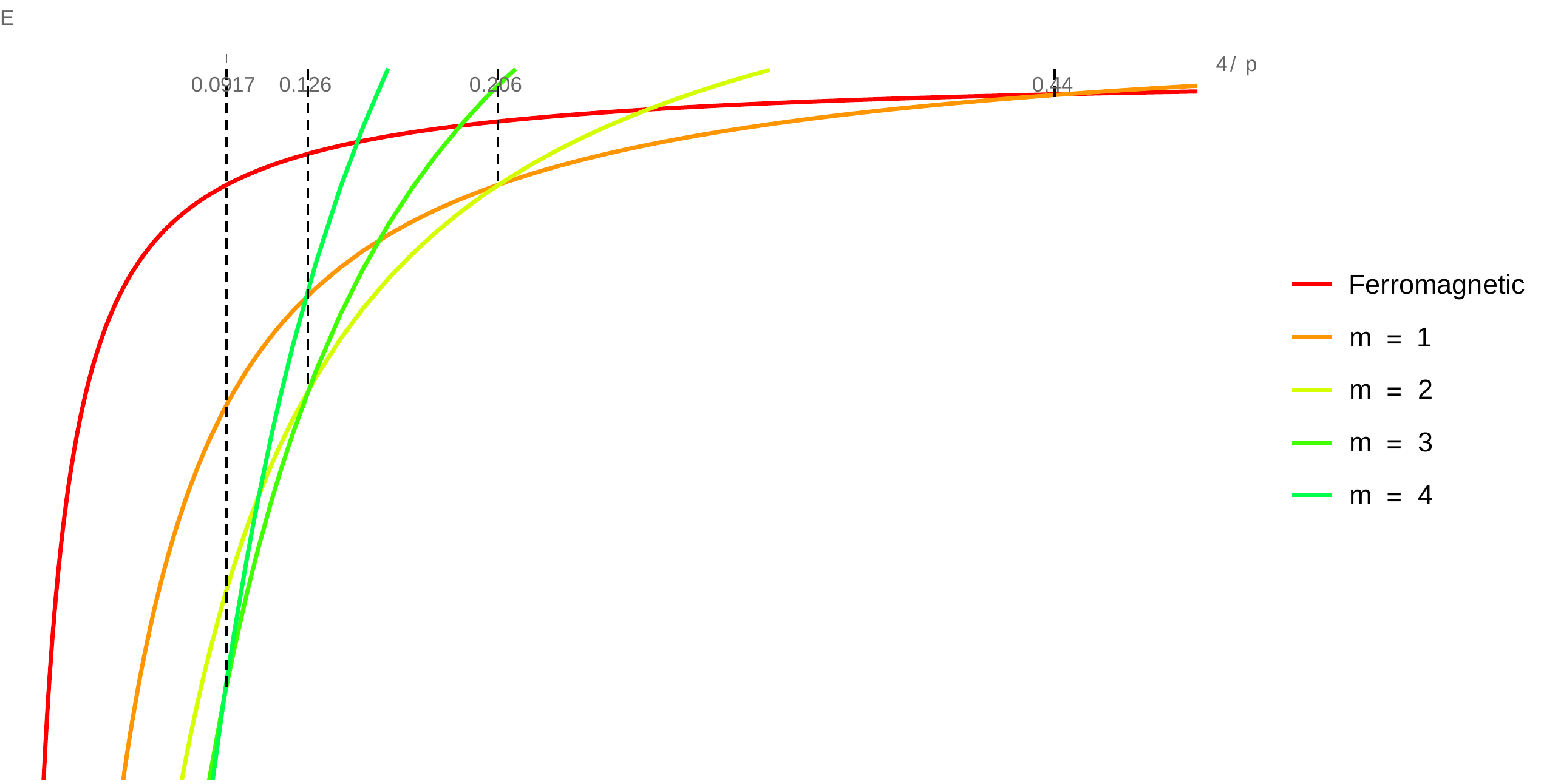}
\caption{
Energies of the clockwise helix for different winding numbers compared with the energy of the ferromagnetic solution. Notice that, as we approach the origin of the horizontal axis $4/p=B/D^2$ from infinity, {\em i.e.} as we decrease the magnetic field or increase the Dzyaloshinskii-Moriya coupling, the vacuum corresponds to solutions with growing winding. 
\label{helixenergyminus}}
\end{figure}

\begin{figure}[H]
\includegraphics[height=8cm]{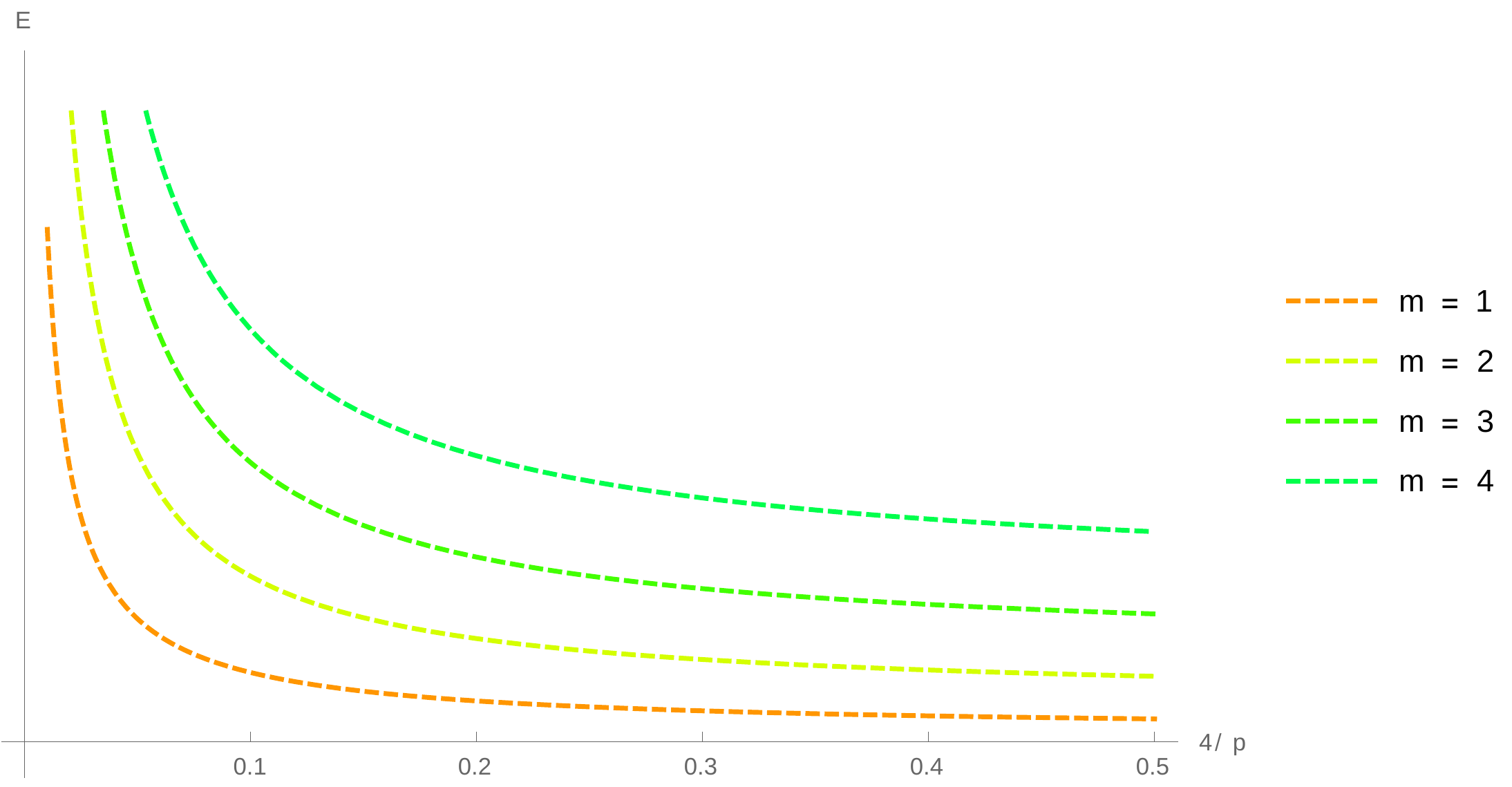}
\caption{
Energies of the counterclockwise helix for different windings. Notice that these solutions have higher energies as compared to the clockwise ones, and that there are no intersections among them. 
\label{helixenergyplus}}
\end{figure}

The on-shell Hamiltonian, when evaluated on the helix solutions takes the form
%
%
\begin{equation}
H^{\sf on-shell}=JL_yL_x
\left(
\left(
\pm \frac{m \pi}{L_x} -1\right)\,p
+
\frac{8 m^2}{L_x^ 2} \,K\!\left(-p_0\right)
\left(
2
\,E\!\left(-p_0\right)
-
K\!\left(-p_0\right)
\right)\right)\,,
\label{eq:onshellhelix}
\end{equation}
where $E(\cdot)$ is the Complete Elliptic Integral of the Second Kind. The important point is that, due to the Dzyaloshinskii-Moriya term, chirality is broken and the two helices have different energy. A plot of the resulting energies for different winding numbers is shown in Figs. \ref{helixenergyminus} and \ref{helixenergyplus}.

\subsection{Viviani solutions}
\label{sec:viviani}
Now we propose a different ansatz, in the form
\begin{eqnarray}
\lambda&=&\lambda_0\,,
\end{eqnarray}
with $\lambda_0$ a constant. When plugged into the equations of motion, we obtain
\begin{eqnarray}
&&X''+\lambda_0 X=0\,,
\nonumber\\
&&\mathcal{Z}''-i p\mathcal{Z}' +
\lambda_0 \mathcal{Z} + i p=0\,,
\nonumber\\
&&X^2+ \mathcal{Z}\mathcal{Z}^*-1=0\,.
\end{eqnarray}
The first equation is solved by an oscillatory solution
\begin{eqnarray}
X &=& X_0\cos\left(\sqrt{\lambda_0}\,x +\delta \right)\,,
\end{eqnarray}
where $X_0$ and $\delta$ are real integration constants. The second equation is solved by
\begin{eqnarray}
{\mathcal{Z}}&=&{\mathcal{Z}}_0e^{i\beta x}-\frac{i p}{\lambda_0}\,,
\end{eqnarray}
where $\mathcal{Z}_0$ is a complex integration constant and we assumed $\lambda_0\neq0$. The wavenumber $\beta$ is given by   $\beta_{\pm}= p/2\pm\sqrt{p^2/4+\lambda_0}$. The last equation then becomes
\begin{equation}
X_0^2\cos^2(\sqrt{\lambda_0}\,x +\delta)+|{\mathcal{Z}}_0|^2+\frac{p^2}{\lambda_0^2}-
\frac{2p}{\lambda_0}{\rm Im}\left[{\mathcal{Z}}_0 e^{i\beta_\pm x}\right]
-1=0\,,
\end{equation}
and writing $\mathcal{Z}_0=|\mathcal{Z}_0|e^{i\varphi}$ we get
\begin{equation}
\frac{X_0^2}2+\frac{X_0^2}2\cos(2\sqrt{\lambda_0}\,x +2\delta)+|{\mathcal{Z}}_0|^2+\frac{p^2}{\lambda_0^2}-
\frac{2p}{\lambda_0}|\mathcal{Z}_0|\sin\left(\beta_\pm x+\varphi\right)
-1=0\,.
\end{equation}
To fulfill this last condition the coefficient of each linearly independent function must cancel. In other words, the following set of equations must be satisfied
\begin{eqnarray}
&&
|\mathcal{Z}_0|^2
+\frac{p^2}{\lambda_0^2}
+\frac{X_0^2}2
-1=0\,,
\nonumber\\
&&
\frac{2p}{\lambda_0}|\mathcal{Z}_0|
=\frac{X_0^2}2\,,
\nonumber\\
&&
\varphi=2\delta+\frac \pi2\,,
\nonumber\\
&&
\beta_\pm= 2\sqrt{\lambda_0}\,.
\end{eqnarray}
The last equation can be satisfied only by $\beta_+$, and determines $\lambda_0=4p^2/9$. The remaining equations are then solved as
\begin{eqnarray}
&&
|\mathcal{Z}_0|
=1-\frac{9}{4p}\,,
\nonumber\\
&&
{X_0^2}=\frac{9}{p}\left(1-\frac{9}{4p}\right)\,,
\nonumber\\
&&
\varphi= 2\delta+\frac \pi2\,.
\end{eqnarray}
Notice that, since we need $|\mathcal{Z}_0|\geq 0$ and ${X_0^2}\geq0$, meaningful solutions exist for $p\geq 9/2$. The $\pm$ sign in $X_0$ resulting from the square root can be reabsorbed by a phase shift $\delta \to\delta+\pi$, so we can fix it to be positive.  The resulting solutions read
\begin{eqnarray}
X &=& 
\frac{3\sqrt{4p\!-\!9}}{2p}
\cos\left(\frac{2p}3\,x +\delta \right)\,,\\
{\cal Z}&=&-\frac{i}{4p}\left(\left(9-4p\right)e^{i\left(\frac{4p}3 x+2\delta\right)}+9\right)\,.
\end{eqnarray}  
These are periodic solutions with period $3\pi/p$, thus the boundary conditions imply $p=3k\pi/L_x$ with $k\in\mathbb{Z}$. For any value of $k$ the solution winds $m=2k$ times around the $x$ axis. Notice that these solutions exist only for discrete values of the parameter $p$. This may imply that they are truly isolated solutions, or that for other values of $p$ their form is not captured by the degrees of freedom of our Ansatz. 

\begin{figure}[ht]
~~~~
\includegraphics[height=8cm]{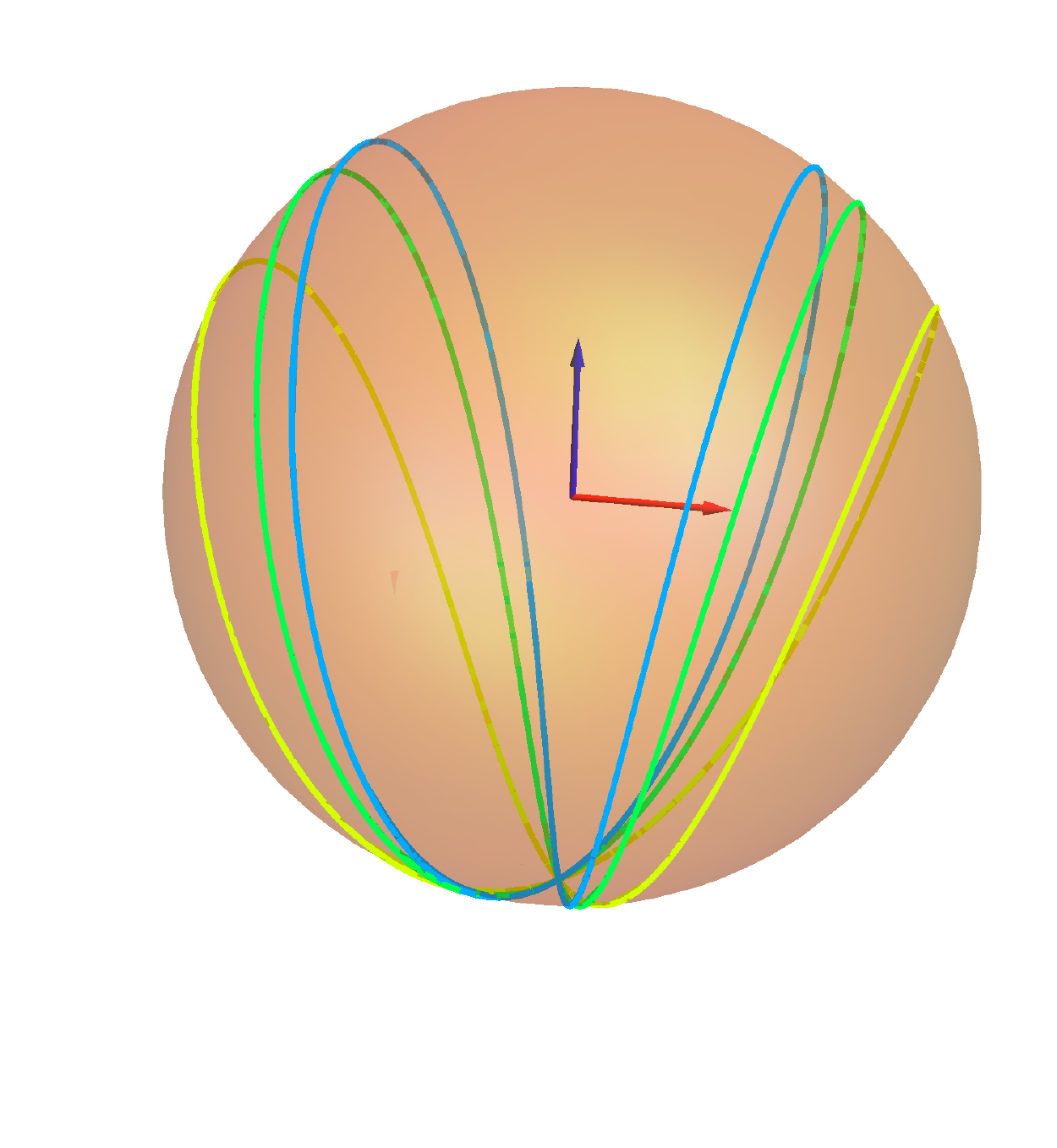}
~~~~~~~~
\includegraphics[height=8cm]{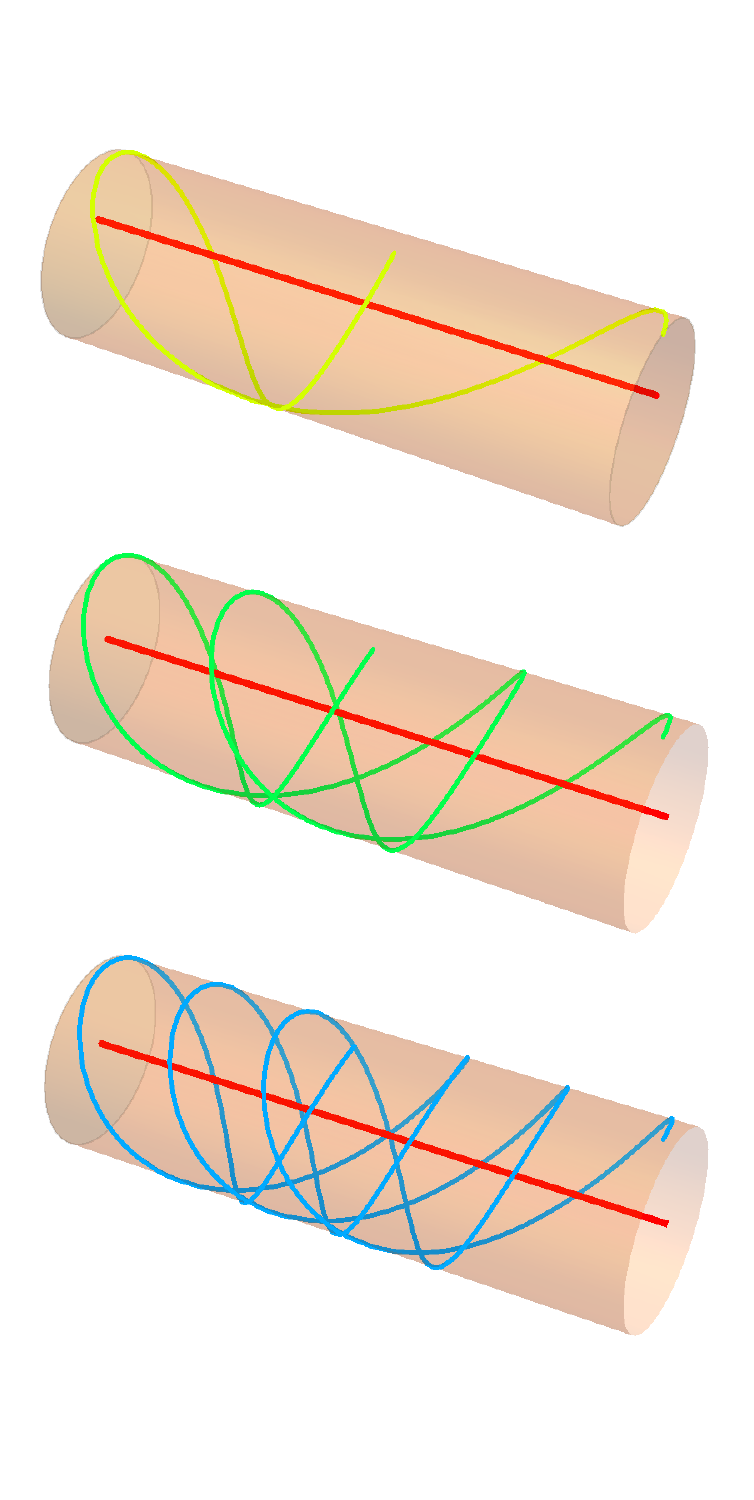}
\caption{The Viviani solutions. It plots a $\infty$-shaped curve on the target sphere (left), oriented according the magnetic field (blue arrow) and perpendicular to the helix axis (red arrow). As it develops in real space as a function of $x$ (right), it turns around the $x$-axis (red line) clockwise. The number $m=2k$ determines the winding, being $2,4,6$ from top to bottom.
\label{viviani}}
\end{figure}

On the target space, these kind of solutions represent the intersection of a circular cylinder of different radii whose axis lies along the $x$ direction with the unit  sphere, with the cylinder and the sphere being tangent at the south pole. The one for which the cylinder has radius $1/2$ is called a {\em Viviani curve} \bibnote{Vincenzo Viviani was a pupil of Evangelista Torricelli, and Galileo Galilei's last disciple and first biographer {\cite{Gattei}}. He studied his famous curve as a solution of an Architectural problem.}, and is a particular case of a class of curves known as {\em Cl\'elias}\bibnote{Named after Clelia Borromeo{\cite{Bardazza2005}} by Luigi Guido Grandi, court mathematician to the Grand Duke of Tuscany{\cite{Agarwal}}, these curves are parametrized by and integer $q$, the value $q=1$ is the Viviani curve{\cite{Gray1998}}}. Plots of the solutions for different values of $k$ are shown in Fig. \ref{viviani}.  The electromagnetic analog allows for an interpretation of the solutions: as the analog particle climbs the sphere from the southern pole pushed by the  electric force, the magnetic force makes the trajectory to curve until it reaches a maximum latitude, and then turns back to the pole. 

\begin{figure}[H]
\centering
\includegraphics[height=8cm]{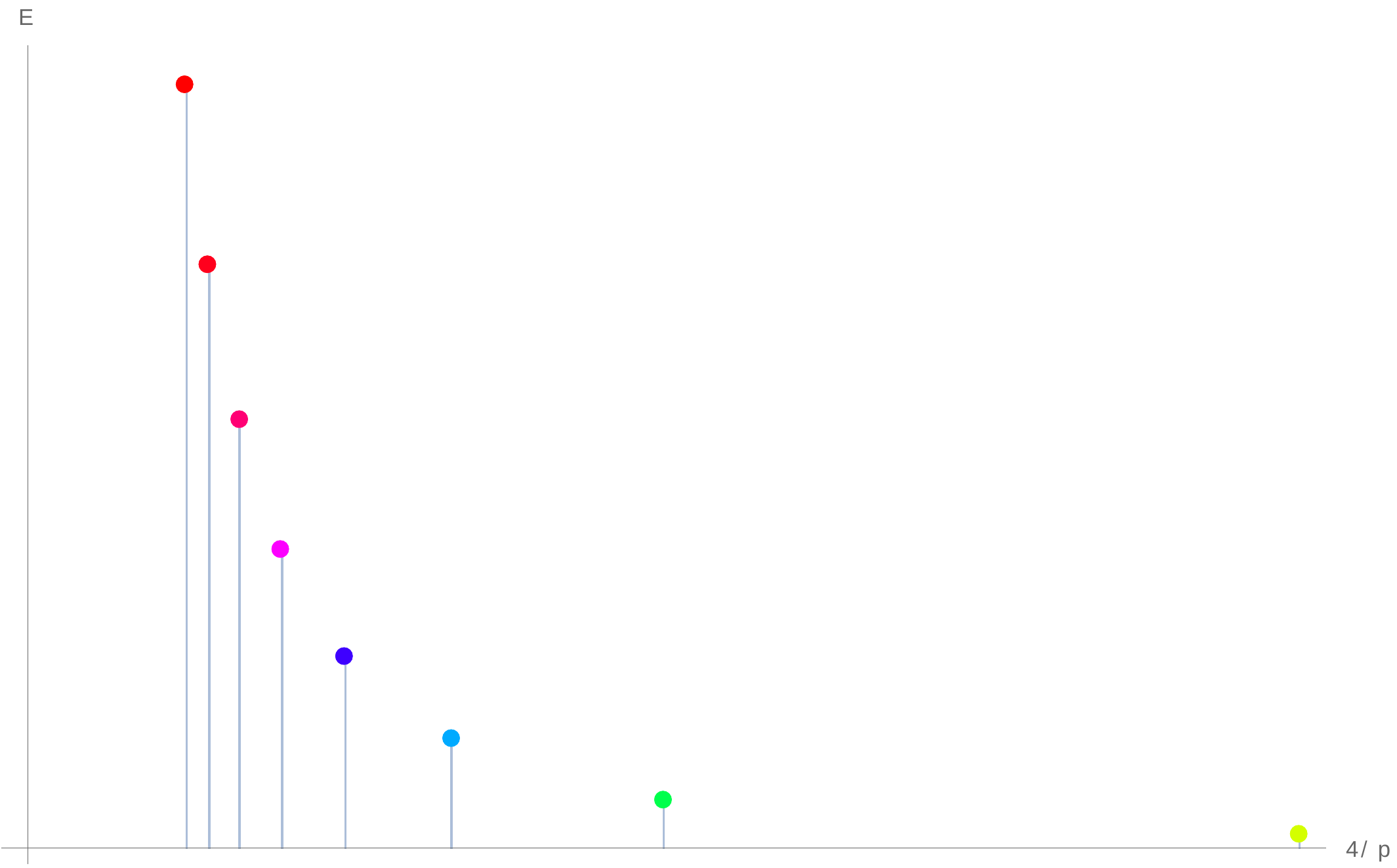}\caption{Values of the energy for the first Viviani solutions. The value of the winding number $m=2k$ increases in steps of $2$ from right to left, the rightmost point corresponding to $m=2$.
\label{vivianienergy}}
\end{figure}

The on-shell Hamiltonian for the Viviani solutions can be easily written by replacing them into the form \eqref{eq:honshell} of the Hamiltonian, obtaining
\begin{equation}
H^{\sf on-shell}=JL_y\left(\frac{2 \pi ^2 }{L_x}k^2+\frac{9 L_x}{8}\right)\,.
\label{eq:onshellviviani}
\end{equation}
A plot of the resulting discrete energies can be seen in Fig. \ref{vivianienergy}.

%

\section{Discussion}
\label{sec:discussion}
\begin{figure}[H]
\includegraphics[height=8cm]{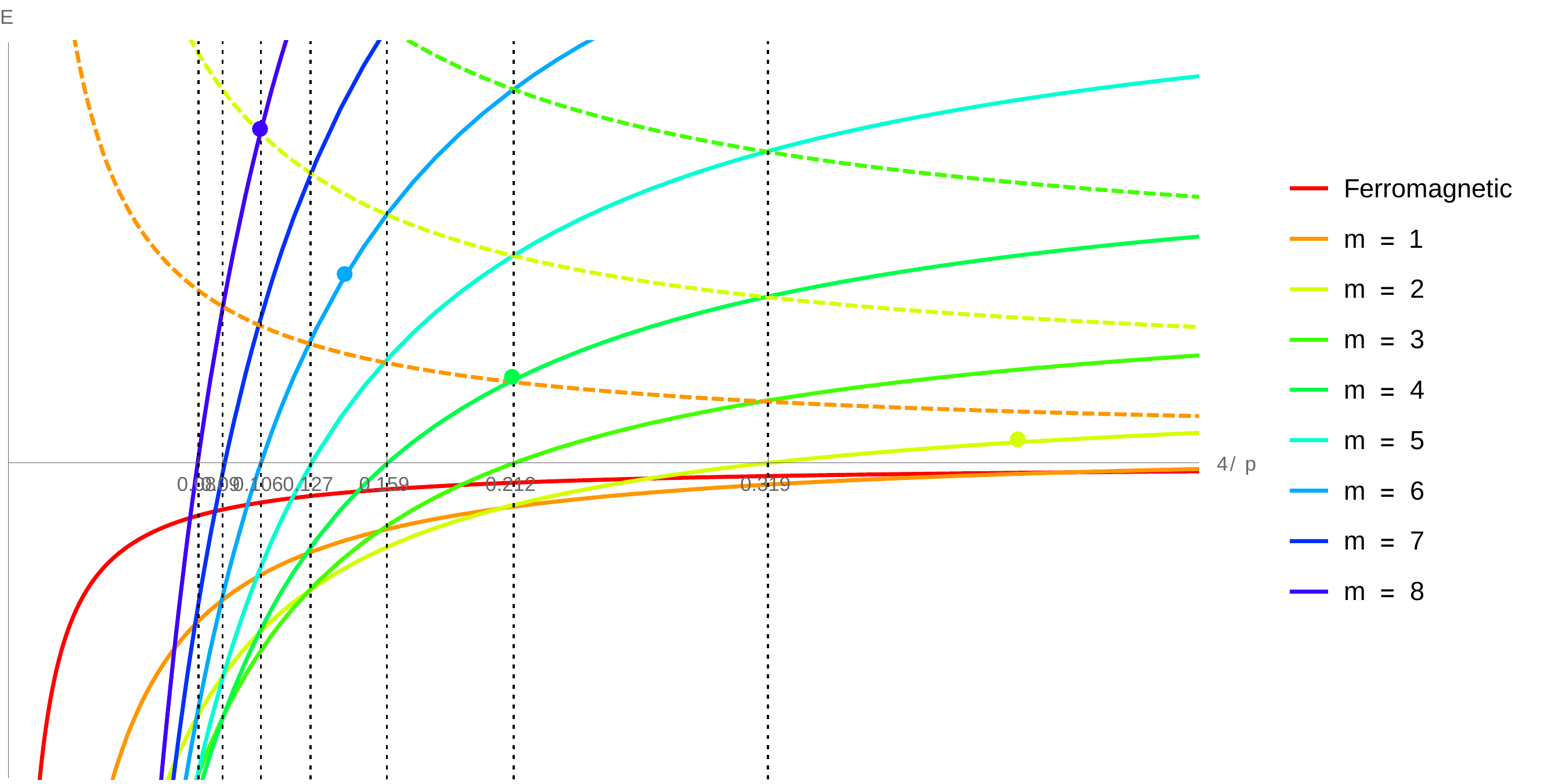}
\caption{
Energies of the helix solutions with both chiralities and the Viviani solutions. The continuous curves correspond to the clockwise helices, the dashed ones to the counterclockwise ones, while the dots represent the Viviani solutions.
Note that the energies of the clockwise and the Viviani solutions always coincide. Moreover, for $m=4, 8$, the energies coincide with those of the counterclockwise helices with $m=1, 2$, respectively. The dotted vertical lines represent the points in the $4/p$ axis at which the spectrum is approximately degenerate. \label{energycompared}}
\end{figure}

When the energies of all the three families are shown in the same plot as in Fig. \ref{energycompared}, an interesting structure appears. This structure includes approximate degeneracies and it can be explained analytically using our formulas above, as follows:
\begin{itemize}
\item The energy of the clockwise helix with winding number $m$ vanishes at values $p_m$ of the parameter $p$ that satisfy numerically $p_m\simeq 2m\pi/L_x$.

In order to prove that analytically, notice that \eqref{eq:K} has the trivial solution $p_0=0$ when the right hand side vanishes. Then, if the right hand side is small enough, we can expand the left hand side in powers of $p_0$ to linear order, and solve it as
\begin{equation}
p_0=
\left(
\frac{ L_x}{m\pi}\right)^2p+\dots\,,
\end{equation}
With this, the on-shell energy \eqref{eq:onshellhelix} can be expanded as
\begin{equation}
H^{\sf on-shell}=JL_y
m\pi
\left(
\frac{2m\pi}{L_x}
\pm p
\right)+\dots\,,
\label{eq:onshellpm}
\end{equation}
which in the case of the clockwise helix ($-$ sign) vanishes at $p_m\simeq 2m\pi/L_x$. 

\item At $p=p_m$, the energy of the clockwise helix with winding $m'\neq m$ coincides approximately with that of the counterclockwise helix with winding $m''=m'-m$.

In order to prove that, in the same approximation as before we write the energy of a helix with winding $m''$ as
\begin{equation}
H^{\sf on-shell}=JL_y
m''\pi
\left(
\frac{2m''\pi}{L_x}
\pm p
\right)+\dots\,,
\end{equation}
Evaluating it at $p_m$ we get
\begin{equation}
H^{\sf on-shell}=\frac{2JL_y\pi^2}{L_x}
m''
\left(
m''
\pm 
m
\right)+\dots\,,
\end{equation}
Now if we put $m''=m'$ with the $-$ sign, we get exactly the same energy as putting $m''=m'-m$ and the $+$ sign, proving what was observed numerically.

\newpage

\item The energy of the Viviani solutions coincides approximately with that of the clockwise helix with the same winding number.

To verify that analytically, we write the energy \eqref{eq:onshellviviani} in terms of the winding $m=2k$ in the form
\begin{equation}
H^{\sf on-shell}=JL_y\,
\frac{m^2\pi^2}{2L_x}+\dots\,.
\label{eq:onshellvivianim}
\end{equation}
where, recalling that Viviani solutions exists only for $p=3m\pi/2L_x$, we discarded a constant term which is small under our present approximation. With such value of $p$ this coincides with the form \eqref{eq:onshellpm} with the $-$ sign, corresponding to a clockwise helix. 

\item Viviani solutions with even $k$ exist at $p\simeq p_m$, with winding $m'=2k$ that is a multiple of $4$. Their energies coincide approximately with both the clockwise helix with winding $m'$ and the counterclockwise helix with winding $m''=m-m'$.

This is easy to check by writing $m=m'-m''$ and then $p_m=2(m'\!-\!m'')\pi/L_x$. Now imposing $p_m=3m'\pi/2L_x$ we get $m'=4m''$.
\end{itemize}

This suggests the following interpretation: at $p\simeq 2m\pi/L_x$ the spectrum becomes approximately degenerate, implying that a new approximate symmetry of the Hamiltonian emerges. 

For $m$ even, the energy of each clockwise solution coincides with that of a counterclockwise solution, implying that the degeneracy is two-fold, corresponding to a discrete $\mathbb{Z}_2$ symmetry.

For $m$ odd, the energy of each clockwise solution coincides with that of a counterclockwise solution as before. But for one particular pair of clockwise and counterclockwise solutions there is a Viviani solution with the same energy. Since this implies a three-fold degeneracy, we can conjecture that at those values of $p$ new solutions could exist, whose energy coincide with that of the remaining pairs of clockwise and counterclockwise solutions. 

Finally, at the points on the $p$ axis where there is a Viviani solution with a winding $m''=2k$ which is not a multiple of $4$, its energy coincide with that of a clockwise solution. This implies that there is a two fold degeneracy, suggesting that new solutions could exist a those points with the same energy as the remaining clockwise solutions.

In consequence, the structure of the energy spectrum suggests the existence of new hypothetical solutions at isolated points of the $p$ axis. These solutions are not captured by the symmetries of our Ansatz, and that is why they were not found in our analysis. A more general Ansatz may provide solutions on the whole $p$ axis that reduce to the isolated Viviani solutions or to the hypothetical new solutions at the particular values of $p$ analyzed above. Since to the moment this conclusion is just a conjecture, further research is needed to determine whether this is the case.

\section{Aknowledgements}
 The authors thank Fabrizio Canfora for useful comments and his collaboration in the early stages on this work. They are also grateful to Gerardo Rossini and Mauricio Sturla for reading the draft and making very useful comments. M.L. appreciates the support of FONDECYT postdoctoral grant 3190873. This work has been funded by the FONDECYT grant 1181047, CONICET grants PIP-2017-1109 and PUE 084 ``B\'usqueda de Nueva F\'\i sica'' and UNLP grant PID-X791. This work was also partially supported by CONICYT
grant PAI80160018.

\newpage
\bibliographystyle{ieeetr}
\bibliography{Mendeley.bib}

\begin{thebibliography}{10}

\bibitem{Dzyaloshinskii1964}
I.~E. Dzyaloshinskii, ``{Theory of helicoidal structures in antiferromagnets.
  I. Nonmetals},'' {\em J. Exptl. Theoret. Phys. (U.S.S.R.)}, vol.~46, no.~4,
  pp.~1420--1437, 1964.

\bibitem{Yablonskii1989}
D.~A. Yablonskii and A.~N. Bogdanov, ``{Thermodynamically stable "vortices" in
  magnetically ordered crystals. The mixed state of magnets},'' {\em Zh. Eksp.
  Teor. Fiz.}, vol.~95, no.~January, pp.~178--182, 1989.

\bibitem{Bogdanov1994}
A.~Bogdanov and A.~Hubert, ``{Thermodynamically stable magnetic vortex states
  in magnetic crystals},'' {\em Journal of Magnetism and Magnetic Materials},
  vol.~138, pp.~255--269, dec 1994.

\bibitem{Bogdanov1999}
A.~Bogdanov and A.~Hubert, ``{The stability of vortex-like structures in
  uniaxial ferromagnets},'' {\em Journal of Magnetism and Magnetic Materials},
  vol.~195, pp.~182--192, apr 1999.

\bibitem{Muhlbauer2009}
S.~Muhlbauer, B.~Binz, F.~Jonietz, C.~Pfleiderer, A.~Rosch, A.~Neubauer,
  R.~Georgii, and P.~Boni, ``{Skyrmion Lattice in a Chiral Magnet},'' {\em
  Science}, vol.~323, pp.~915--919, feb 2009.

\bibitem{Ishikawa1984}
Y.~Ishikawa and M.~Arai, ``{Magnetic Phase Diagram of MnSi near Critical
  Temperature Studied by Neutron Small Angle Scattering},'' {\em Journal of the
  Physical Society of Japan}, vol.~53, no.~8, pp.~2726--2733, 1984.

\bibitem{Lebech1995}
B.~Lebech, P.~Harris, J.~{Skov Pedersen}, K.~Mortensen, C.~I. Gregory, N.~R.
  Bernhoeft, M.~Jermy, and S.~A. Brown, ``{Magnetic phase diagram of MnSi},''
  {\em Journal of Magnetism and Magnetic Materials}, vol.~140-144,
  pp.~119--120, feb 1995.

\bibitem{Shibata2013}
K.~Shibata, X.~Z. Yu, T.~Hara, D.~Morikawa, N.~Kanazawa, K.~Kimoto,
  S.~Ishiwata, Y.~Matsui, and Y.~Tokura, ``{Towards control of the size and
  helicity of skyrmions in helimagnetic alloys by spin–orbit coupling},''
  {\em Nature Nanotechnology}, vol.~8, pp.~723--728, oct 2013.

\bibitem{Lebech1989}
B.~Lebech, J.~Bernhard, and T.~Freltoft, ``{Magnetic structures of cubic FeGe
  studied by small-angle neutron scattering},'' {\em Journal of Physics:
  Condensed Matter}, vol.~1, pp.~6105--6122, sep 1989.

\bibitem{Uchida2008}
M.~Uchida, N.~Nagaosa, J.~P. He, Y.~Kaneko, S.~Iguchi, Y.~Matsui, and
  Y.~Tokura, ``{Topological spin textures in the helimagnet FeGe},'' {\em
  Physical Review B}, vol.~77, p.~184402, may 2008.

\bibitem{Yu2011}
X.~Z. Yu, N.~Kanazawa, Y.~Onose, K.~Kimoto, W.~Z. Zhang, S.~Ishiwata,
  Y.~Matsui, and Y.~Tokura, ``{Near room-temperature formation of a skyrmion
  crystal in thin-films of the helimagnet FeGe},'' {\em Nature Materials},
  vol.~10, pp.~106--109, feb 2011.

\bibitem{Wilhelm2011}
H.~Wilhelm, M.~Baenitz, M.~Schmidt, U.~K. R{\"{o}}{\ss}ler, A.~A. Leonov, and
  A.~N. Bogdanov, ``{Precursor Phenomena at the Magnetic Ordering of the Cubic
  Helimagnet FeGe},'' {\em Physical Review Letters}, vol.~107, p.~127203, sep
  2011.

\bibitem{Beille1983}
J.~Beille, J.~Voiron, and M.~Roth, ``{Long period helimagnetism in the cubic
  B20 Fe$_x$Co$_{1-x}$Si and Co$_x$Mn$_{1-x}$Si alloys},'' {\em Solid State
  Communications}, vol.~47, no.~5, pp.~399--402, 1983.

\bibitem{Grigoriev2007}
S.~V. Grigoriev, V.~A. Dyadkin, D.~Menzel, J.~Schoenes, Y.~O. Chetverikov,
  A.~I. Okorokov, H.~Eckerlebe, and S.~V. Maleyev, ``{Magnetic structure of
  Fe$_{1-x}$Co$_x$Si in a magnetic field studied via small-angle polarized
  neutron diffraction},'' {\em Physical Review B}, vol.~76, no.~22, p.~224424,
  2007.

\bibitem{Grigoriev2009}
S.~V. Grigoriev, D.~Chernyshov, V.~A. Dyadkin, V.~Dmitriev, S.~V. Maleyev,
  E.~V. Moskvin, D.~Menzel, J.~Schoenes, and H.~Eckerlebe, ``{Crystal
  Handedness and Spin Helix Chirality in Fe$_{1-x}$Co$_x$Si},'' {\em Physical
  Review Letters}, vol.~102, no.~3, p.~37204, 2009.

\bibitem{Onose2005}
Y.~Onose, N.~Takeshita, C.~Terakura, H.~Takagi, and Y.~Tokura, ``{Doping
  dependence of transport properties in Fe$_{1-x}$Co$_x$Si},'' {\em Physical
  Review B}, vol.~72, no.~22, p.~224431, 2005.

\bibitem{Kezsmarki2015}
I.~K{\'{e}}zsm{\'{a}}rki, S.~Bord{\'{a}}cs, P.~Milde, E.~Neuber, L.~M. Eng,
  J.~S. White, H.~M. R{\o}nnow, C.~D. Dewhurst, M.~Mochizuki, K.~Yanai,
  H.~Nakamura, D.~Ehlers, V.~Tsurkan, and A.~Loidl, ``{N{\'{e}}el-type skyrmion
  lattice with confined orientation in the polar magnetic semiconductor{\~{}}
  GaV$_4$S$_8$},'' {\em Nature Materials}, vol.~14, pp.~1116--1122, nov 2015.

\bibitem{Seki2012}
S.~Seki, X.~Z. Yu, S.~Ishiwata, and Y.~Tokura, ``{Observation of skyrmions in a
  multiferroic material.},'' {\em Science (New York, N.Y.)}, vol.~336,
  no.~6078, pp.~198--201, 2012.

\bibitem{Adams2012}
T.~Adams, A.~Chacon, M.~Wagner, A.~Bauer, G.~Brandl, B.~Pedersen, H.~Berger,
  P.~Lemmens, and C.~Pfleiderer, ``{Long-Wavelength Helimagnetic Order and
  Skyrmion Lattice Phase in Cu$_2$OSeO$_3$},'' {\em Physical Review Letters},
  vol.~108, p.~237204, jun 2012.

\bibitem{Baskaran2011a}
G.~Baskaran, ``{Possibility of Skyrmion Superconductivity in Doped
  Antiferromagnet K$_2$Fe$_4$Se$_5$},'' {\em arXiv:1108.3562}, aug 2011.

\bibitem{Yi2009}
S.~D. Yi, S.~Onoda, N.~Nagaosa, and J.~H. Han, ``{Skyrmions and anomalous Hall
  effect in a Dzyaloshinskii-Moriya spiral magnet},'' {\em Physical Review B},
  vol.~80, no.~5, p.~54416, 2009.

\bibitem{Han2010}
J.~H. Han, J.~Zang, Z.~Yang, J.-H. Park, and N.~Nagaosa, ``{Skyrmion lattice in
  a two-dimensional chiral magnet},'' {\em Physical Review B}, vol.~82,
  p.~94429, sep 2010.

\bibitem{Grandi2018a}
N.~Grandi and M.~Sturla, ``{Approximate analytic expression for the Skyrmions
  crystal},'' {\em International Journal of Modern Physics B}, vol.~32, no.~2,
  2018.

\bibitem{Nogueira2018}
F.~S. Nogueira, I.~Eremin, F.~Katmis, J.~S. Moodera, J.~van~den Brink, and
  V.~P. Kravchuk, ``{Fluctuation-induced N{\'{e}}el and Bloch skyrmions at
  topological insulator surfaces},'' {\em Physical Review B}, vol.~98,
  p.~060401, aug 2018.

\bibitem{Note1}
Vincenzo Viviani was a pupil of Evangelista Torricelli, and Galileo Galilei's
  last disciple and first biographer {\cite {Gattei}}. He studied his famous
  curve as a solution of an Architectural problem.

\bibitem{Note2}
Named after Clelia Borromeo{\cite {Bardazza2005}} by Luigi Guido Grandi, court
  mathematician to the Grand Duke of Tuscany{\cite {Agarwal}}, these curves are
  parametrized by and integer $q$, the value $q=1$ is the Viviani curve{\cite
  {Gray1998}}.

\bibitem{Gattei}
S.~Gattei, {\em {On the life of Galileo : Viviani's Historical account and
  other early biographies}}.
\newblock Princeton University Press, 2019.

\bibitem{Bardazza2005}
A.~M. {Serralunga Bardazza}, {\em {Clelia Grillo Borromeo Arese: vicende
  private e pubbliche virt{\`{u}} di una celebre nobildonna nell'Italia del
  Settecento}}.
\newblock Eventi {\&} Progetti, 2005.

\bibitem{Agarwal}
R.~P. Agarwal and S.~K. Sen, {\em {Creators of mathematical and computational
  sciences}}.
\newblock Cham: Springer International Publishing, 2014.

\bibitem{Gray1998}
A.~Gray, {\em {Modern differential geometry of curves and surfaces with
  Mathematica}}.
\newblock CRC Press, 1998.

\end{thebibliography}
\end{document}